\documentclass{article}

\PassOptionsToPackage{numbers, compress}{natbib}

\usepackage[preprint]{neurips_2024}

\usepackage[utf8]{inputenc} 
\usepackage[T1]{fontenc}    
\usepackage{hyperref}       
\usepackage{url}            
\usepackage{booktabs}       
\usepackage{amsfonts}       
\usepackage{nicefrac}       
\usepackage{microtype}      
\usepackage{xcolor}         
\usepackage{times}
\usepackage{amsmath}
\usepackage{mathabx}
\usepackage{multirow}
\usepackage{graphicx}
\usepackage{xspace}
\usepackage{float}
\usepackage{enumitem}
\usepackage[ruled]{algorithm2e}
\SetAlFnt{\small}
\SetAlCapFnt{\small}
\SetAlCapNameFnt{\small}
\usepackage{algpseudocode}
\usepackage{tikz}
\usepackage{cancel}
\usepackage{caption}
\usepackage{subcaption}
\usepackage{makecell}
\usepackage{tablefootnote}

\usepackage{xcolor}

\newcommand{\redtext}[1]{{\small \color{red} #1}}

\newcommand{\bluetext}[1]{{\small \color{blue} #1}}

\definecolor{purple}{rgb}{0.5,0,1}
\definecolor{dcyan}{rgb}{0.2,0.6,0.5}
\definecolor{light-gray}{gray}{0.95} 
\definecolor{darkgreen}{RGB}{0,140,0}
\definecolor{darkred}{RGB}{200,0,0}
\definecolor{lightgreen}{RGB}{189,252,192}
\definecolor{lightred}{RGB}{255,205,212}
\definecolor{lightyellow}{RGB}{255,240,160}
\definecolor{lightblue}{RGB}{195,221,255}
\definecolor{lightpurple}{RGB}{232,209,255}

\usepackage{pifont}

\usepackage{array}
\newcolumntype{?}[1]{!{\vrule width #1}}

\title{\textsc{Revision Matters}: \\ Generative Design Guided by Revision Edits}

\author{%
  Tao Li \\
  Google DeepMind\\
  Mountain View, CA, USA \\
  \texttt{tlinlp@google.com} \\
  \And
  Chin-Yi Cheng \\
  Google DeepMind\\
  Mountain View, CA, USA \\
  \texttt{cchinyi@google.com} \\
  \And
  Amber Xie \\
  Stanford University\\
  Stanford, CA, USA \\
  \texttt{amberxie88@gmail.com} \\
  \And
  Gang Li \\
  Google DeepMind\\
  Mountain View, CA, USA \\
  \texttt{leebird@google.com} \\
  \And
  Yang Li \\
  Google DeepMind\\
  Mountain View, CA, USA \\
  \texttt{liyang@google.com} \\
}

\begin{document}

\maketitle


\begin{abstract}
  Layout design, such as user interface or graphical layout in general, is fundamentally an iterative revision process. Through revising a design repeatedly, the designer converges on an ideal layout.
  In this paper, we investigate how revision edits from human designer can benefit a multimodal generative model.
  To do so, we curate an expert dataset that traces how human designers iteratively edit and improve a layout generation with a prompted language goal.
  Based on such data, we explore various supervised fine-tuning task setups on top of a Gemini multimodal backbone, a large multimodal model.
  Our results show that human revision plays a critical role in iterative layout refinement.
  While being noisy, expert revision edits lead our model to a surprisingly strong design FID score $\small\sim10$ which is close to human performance ($\small\sim6$).
  In contrast, self-revisions that fully rely on model's own judgement, lead to an echo chamber that prevents iterative improvement, and sometimes leads to generative degradation.
  Fortunately, we found that providing human guidance plays at early stage plays a critical role in final generation.
  In such human-in-the-loop scenario, our work paves the way for iterative design revision based on pre-trained large multimodal models.
  
\end{abstract}

\section{Introduction} \label{sec:intro}


The pursuit of ideal layout design, whether for user interfaces or graphical elements, is fundamentally iterative. Designers meticulously refine their creations through cycles of revision, gradually converging towards a solution that balances aesthetics and functionality. While recent advances in multimodal generative models have enabled automated design generation from textual prompts~\citep{cheng2023play, chen2022pali}, these models often ignore the iterative guidance implied in edits made by human designers, thus fall short of capturing the nuances and iterative nature inherent to human design.

Human revision edits, as a sequence of layout state changes, are often noisy and extend beyond simply, localized changes. For instance, a designer might explore the visual appearances of widgets/elements, lumping them on the working layout, and only keep the favored few. Such attempts can further be interleaved by operations that organize scattered objects for unified shapes and aligned positions. This extended and complex nature of revisions makes it difficult to directly translate them into clear instructions for a generative model, or preference pairs for reward-based models~\cite{xie2024leveraging}.

In this paper, we studies the potential of \emph{directly} using human revision edits to guide and enhance multimodal generative models in layout design.
With a strong pre-trained Gemini backbone~\cite{team2023gemini}, we focus on the impact of various setups that are immediately available to supervised fine-tuning scenario.
Generally, We hypothesize that the iterative process undertaken by human designers, while being noisy and lengthy, encapsulates valuable knowledge that can be harnessed to improve the generated layouts.
To investigate this, we curate a novel dataset, namely \textsc{Rare+}, capturing the iterative design process. This dataset meticulously traces how expert designers revise and refine layouts generated from a specific language prompt, offering a rich source of insights into the evolution of a design.

Using this unique dataset, we explore various supervised fine-tuning setups for a Gemini backbone, a large multimodal model~\citep{team2023gemini}.
We demonstrate that incorporating human revision edits substantially surpasses models that ignored it.
Despite the inherent noise in human edits, our best model variant achieves a remarkable design FID score\footnote{Lower the score, better the performance.}~\citep{heusel2017gans} of ${\small \sim}$10, closely approaching human performance (FID ${\small \sim}$6).
In contrast, the baseline, a latent diffusion model developed for the same problem~\citep{xie2024leveraging}, achieves ${\small \sim}$30.

Furthermore, our research investigates the emergence of self-improvement capabilities within our design generator.
During repeatedly self-feeding at inference time, models quickly exhibit a self-repeating behavior pattern, essentially reaching a tight echo chamber~\cite{sharma2024generative}.
To remedy in this case, we found that injecting human guidance at early stage can substantially outperform models that fully rely on themselves.
This highlights the critical role of human guidance in such iterative design generation task.

In summary, our main contributions are:
\begin{itemize}
    \item We introduce a novel dataset \textsc{Rare+} capturing the iterative nature of human design revisions.
    \item We demonstrate the substantial benefits of incorporating revision edits in training multimodal generative models for layout design.
    \item We study the impact of human revision edits and compare them against model-generated ones in a auto-regressive manner, showing that human revisions are far better than model generated ones in generation guidance, and highlighting the potential of closing this gap.
\end{itemize}

\section{Related Works}

Incorporating human priors has proven effective in guiding language models \citep{liu2023chain, ouyang2022training}, text-to-image models \citep{lee2023aligning, fan2023dpok, black2023training}, and robot behaviors \citep{hejna2022fewshot, lee2021pebble}. These priors primarily manifest in two forms.
\begin{itemize}
    \item Dataset curation: finetuning models on high-quality, human-aligned datasets \citep{ouyang2022training} instills desirable biases;
    \item Reward modeling: Leveraging pre-trained foundation models such as Stable Diffusion \citep{rombach2022highresolution, jain2022vectorfusion} or R3M \citep{nair2022r3m, adeniji2023language} allows for extracting implicit reward functions. Reward functions can also be learned directly from human feedback provides explicit guidance for model alignment \citep{xu2023imagereward, lee2023aligning, bai2022training, stiennon2022learning}.
\end{itemize}

\paragraph{Learning from Human Feedback} Reinforcement learning from human feedback (RLHF) offers a popular two-stage process: training a reward model and optimizing a reinforcement learning objective. This method has demonstrated success in various domains, including language modeling \citep{casper2023open} and text-to-image generation \citep{lee2023aligning, black2023training}.
However, reinforcement learning-free approaches also exist. Supervised finetuning on curated datasets \citep{ouyang2022training} represents a more direct method, while recent advances introduce innovative supervised objectives, such as Chain of Hindsight \citep{liu2023chain} and Direct Policy Optimization \citep{rafailov2023direct}, for aligning large language models (LLMs).

\paragraph{Layout Generation} Vector graphic formats facilitate easy editing and downstream use of generated layouts, making this domain well-suited for collecting human revisions. Recent studies utilize sequential models like Transformers \citep{transformer} to generate layout elements as sequences \citep{gupta2021layouttransformer, vtn, blt, kikuchi2021constrained}. LayoutDM and PLay \citep{inoue2023layoutdm, cheng2023play} demonstrate conditional layout generation capabilities using diffusion models. Two recent works \citep{lin2024layoutprompter, lin2023parse} leverages LLMs for few shot layout generation, but they either require specific selection and ranking process for examples or only use LLMs to parse the input. RARE \citep{xie2024leveraging} was the first attempt to improve UI layout design quality using reward models based human revisions.  

\paragraph{Our Takes} While preferences, such as \citep{bai2022training, liu2023chain, lee2023aligning, stiennon2022learning}, dominate generative model research, what type of feedback to use remains an open question. Closely in line of our work, \citet{xie2024leveraging} found that the layout design generation process from human editing can be fuzzy and extended (hundreds of steps).
Directly extracting or modeling a binary preference reward ended up to be challenging.
Fundamentally, in the layout generation task, we care about if the model can generate the ideal design based on a given starting state.
To this end, we take a step back and look at the direct impact of curated human data on a powerful multimodal model backbone (i.e., Gemini~\citep{team2023gemini}).

\section{Task Formulation}

Given an initial user interface (UI) layout design $S_0$ and a language prompt $T$ that describes the functionality that the UI should achieve, our goal is to generate an improved final version $S_n$ that matches the functionality description and meets designers' aesthetic judgement. From $S_0$ to $S_n$, the human designer makes $n$ revision edits, which each results in an intermediate layout state $S_i$ where $i\in(0, n]$.
In general, each $S_i$ can be represented by an image screenshot $I_i$ and/or a sequence of design code $C_i$.
The choice of using which of these representations depends on the underlying model capabilities, such as the context length and interleaved modalities.

\paragraph{Discussions} There are other closely related task formations in the image-to-text and text-to-image works.
In this paper, we focus on the conditional setup that generates the final design code $C_n$ conditioned on a given initial state $S_0$ and the language prompt $T$, which differs from stateless generation such as $T \rightarrow C_n$ or $T \rightarrow I_n$.

\subsection{\textsc{Rare}+ Dataset}

\begin{figure*}[t]
  \centering
   \includegraphics[width=0.75\linewidth]{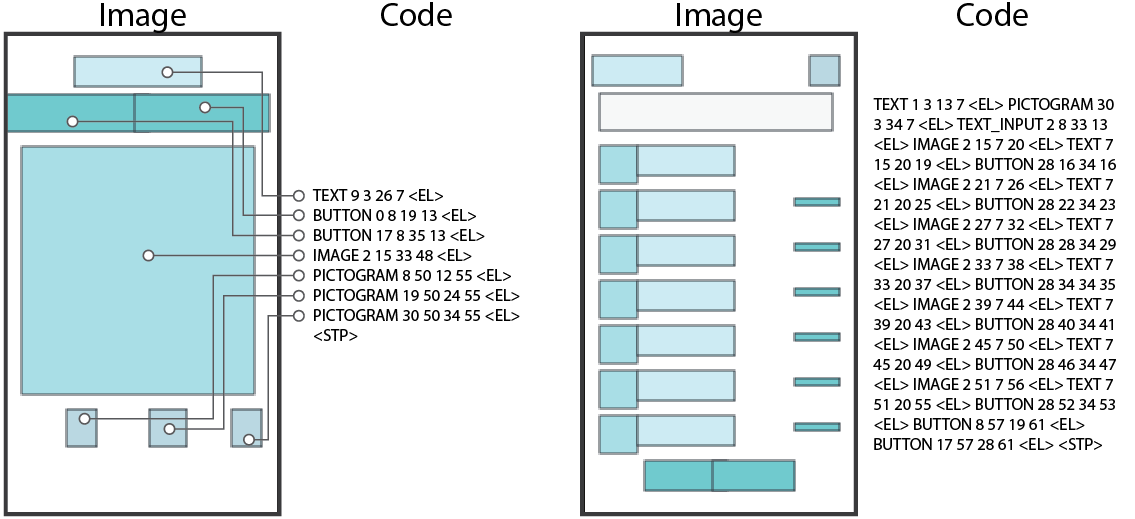}

   \caption{Examples of state (i.e., image-code pair) in \textsc{Rare+} dataset.
   Refer to Appx.~\ref{sec:color_legend} for color legends.}
   \label{fig:image_code_pair}
\end{figure*}

Our goal is to study how human revision edits impact generative model. Previously, \textsc{Rare}~\citep{xie2024leveraging} collected a small scale data based on \textsc{PLay}~\citep{cheng2023play} with human revisions. We follow the same procedure for data collection. In particular, we based our data collection on Figma\footnote{\url{https://www.figma.com/}}, a commonly used tools for the design community. We developed a Figma plugin to render mock-ups generated by PLay~\citep{cheng2023play} and use the associated text condition of each UI layout as the language prompt for the intended functionality of the design. This design of data collection is particularly targeted on the scenario of improving the outcome of a layout generative model. With the plugin, we enable design elements and text boxes from Material 3 design library\footnote{\url{https://m3.material.io/}} in our data collection.
We use the plugin to record all operations performed by the designer and the resultant layout states for each operation.

We give layouts generated by PLay to 4 professional UI designers and ask them to modify the layout to reach an aesthetic and coherent point.
Possible design operations include fixing misaligned elements and changing the format of the layout based on the text prompt (i.e., the text description of the design goal).

Our data collection based on the above process yields a 5 times larger and a cleaner version of the dataset compared to the original \textsc{Rare}, including $5,500$ training examples and $742$ test instances after heuristic filtering.
Fig.~\ref{fig:image_code_pair} shows examples from the dataset where each revision state in a design is a image-code pair. Each design example is essentially a sequence of revision states, representing a human revision process.

\paragraph{Noisy and lengthy edits}
As mentioned in Sec.~\ref{sec:intro}, human revision edits are fuzzy and extended.
During the process, designers often perform experimental changes to gain a look and feel about the design, and then later revert them.
Such a revision pattern can spread in several hundreds of edits.
In Tab.~\ref{tab:rev_steps}, we quantitatively measure the fluctuation in design affinity across the edit operations with respect to the final design.
Specifically, we quantized each edit trajectory into 5 buckets (stages) and measure the FID score of each bucket compared to the ground truth (the final design).
Clearly, design quality does not necessarily follow a monotonous improving pattern.

\begin{table}[h]
  \centering
  \scalebox{1.0}{
  \begin{tabular}{l|c|c|c|c|c}
    \toprule
    Step & $S_0$ & $S_{i25\%}$ & $S_{i50\%}$ & $S_{i75\%}$ & $S_n$ \\
    \midrule
    FID & 47.2 & 63.3 & 37.8 & 21.2 & 6.5 \\
    \bottomrule
  \end{tabular}
  }
  \caption{\small{Along a revision process, human edits are not necessarily 
  monotonously improving a design towards the final outcome.}}
  \label{tab:rev_steps}
\end{table}



\subsection{Modeling} \label{sec:modeling}
\begin{figure*}[h]
  \centering
   \includegraphics[width=0.85\linewidth]{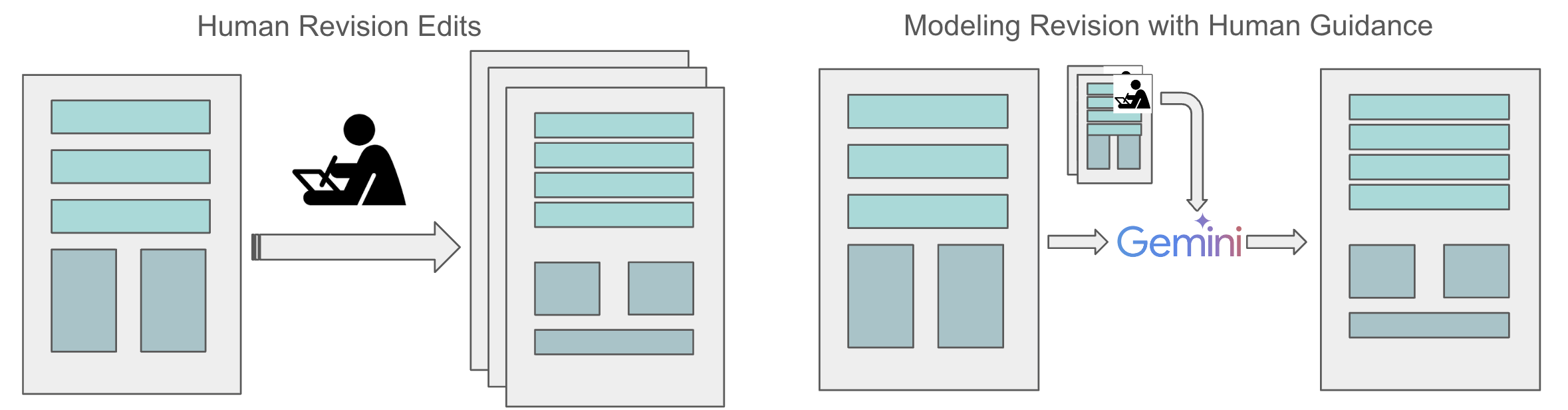}

   \caption{Modeling overview}
   \label{fig:modeling_overview}
\end{figure*}

Our modeling focuses on the layout design transitioning from a prior state to a later state, which essentially takes a given initial state or a human revised one as input, and outputs an improved version of the design, as shown in Fig.~\ref{fig:modeling_overview}.
In this section, we introduce a variation of modeling designs centered on such a setting.

\paragraph{Considerations} Our backbone multimodal model is a pretrained Gemini-1.5~\citep{team2023gemini} with observed general reasoning ability.
Therefore, we retain the architecture untouched, and focus on ablations of input and output data when designing our models.
After all, recall that in Sec.~\ref{sec:intro} we want to see the impact of revision edits in layout generation.
At a first glance, modeling human revision trajectory seems like a good initial point. However, due to the lengthy and noisy edit operations, it is difficult and tricky to define a reliable reward/preference between edits~\citep{xie2024leveraging}.
Therefore, we focus on the jump from a given initial layout to the final design, as the final design can more reliably resemble a desired layout to achieve, and how it works with given human edits as an extra guidance.

Specifically, our modeling paradigm focuses on the image-to-text setup~\cite{chen2022pali} where the input consists of multimodal features and the output is pure text. In our layout generation task, we aim to generate design code, a form of pure text, which will be separately rendered to a screenshot, and is then compared with human generated image designs.
We will use the existing auto-regressive training loss of Gemini on the design code generation.
To evaluate rendered layout images, we use FID scores~\citep{heusel2017gans}, which have been a common metric for visual quality of generative results.

\paragraph{Direct model}
The first model we consider, as a baseline, takes any layout state in the revision sequence as the input and directly generates the final design, for instance, starting with $S_0$.
\begin{align}
    C_n \sim f_\theta(C_0, I_0; T) \label{eq:direct_model}
\end{align}
where $f_\theta$ is the underlying generative model to train. In addition to Eq.~\ref{eq:direct_model}, we also consider the variant that takes a randomly sampled intermediate state from the revision sequence as the input: $C_n \sim f_\theta(C_i, I_i, T)$.

\paragraph{Hop model}
Instead of treating revision edits as a one-hop operation, we can implicitly model the iterative editing process via individual hops:
\begin{align}
    C_j \sim f_\theta(C_i, I_i, T)
\end{align}
where $0\leq i<j \leq n$.
Recall that human revision sequence can be fuzzy and lengthy (hundreds of steps).
Our assumption is that the immediate next step does not necessarily improve the design quality, but over the long run it should statistically.
Therefore, we explore two different sampling strategies on the $(i,j)$ pair:
\begin{enumerate}[nosep]
    \item \emph{j-then-i}: firstly sample $j$ from a Gaussian centered on the 90 percentile of $[0,n]$ and then uniformly sample $i\sim[0,j)$; or
    \item \emph{quantized}: quantize the entire edit sequence into 5 equal buckets (revision stages) and sample $(i,j)$ pair from any different two buckets.
\end{enumerate}
We repeat such sampling for $10$ times, leading to a training set of $55,000$.

\paragraph{Single revision model}
In this case, we want to examine whether adding a human revision edit helps predicting the final state.
\begin{align}
    C_n \sim f_\theta(C_0, I_0, C_i; T) \label{eq:one_step_model}
\end{align}
The hypothesis is that $S_i$ carries directional edit guidance to induce the learning from $S_0$ to $S_i$ and to $S_n$.
We sample $S_i$ uniformly from the intermediate layouts of the human revision trajectory.

\paragraph{Multi-revision model}
In this case, we incorporate more revision context for predicting the final state.
Since human revision edits can be noisy and lengthy, we sample a subset of layout states from the edit sequence, 
as a richer form of guidance.
\begin{align}
    C_n \sim f_\theta(C_0, I_0, \mathcal{C}; T)
\end{align}
where we use the design code sequences $\mathcal{C}$ to represent the sampled revision subset.
Note that we do not use the image tokens associated with $\mathcal{C}$ to have a reasonably context window size.
In practice, the size of $\mathcal{C}$ is uniform in $[0, 20]$.
When the size is $0$, it essentially falls back to single revision model.

\section{Experiments} \label{sec:experiments}

In this section, we focus on evaluating our proposed model setups in Sec.~\ref{sec:modeling}.
In Sec.~\ref{sec:example_construction}, we discuss how our multimodal inputs are constructed.
Sec.~\ref{sec:train_eval} presents learning and inference details.
In Sec.~\ref{sec:direct_inf}, we report performances that ignores human revision at inference time.
Moreover, Sec.~\ref{sec:human_guided} reports the impact of human guidance at test time.

\subsection{Multimodal Example Construction} \label{sec:example_construction}

For models discussed in Sec.~\ref{sec:modeling}, we formulate our input example as a sequence of tokens, including both text and visual ones.
The input state representations (i.e., design code $C$ and image $I$) are interleaved by a natural language instruction.
We use such a language instruction to hint the pretrained model about image and text co-reference.
During supervised fine-tuning, the input sequence in Tab.~\ref{tab:multimodal_seq} is frozen, and the training only supervises on the output text tokens.

\begin{table}[h]
  \centering
  \scalebox{0.97}{
  \begin{tabular}{l|p{0.7\linewidth}}
    \toprule
    Model & Input sequence \\
    \midrule
    Direct/Hop & Your are improving the layout design of an app. [\bluetext{\texttt{T}}] The initial layout is: [\bluetext{\texttt{C$_{0/i}$}}] Now, improve the layout based on the initial layout's screenshot: [\redtext{\texttt{I$_{0/i}$}}] \\
    \midrule
    Single/Multi-revision & Your are improving the layout design of an app. [\bluetext{\texttt{T}}] The initial layout is: [\bluetext{\texttt{C$_{0}$}}] You made some edits to the initial layout: [\bluetext{\texttt{C$_{i}$}}/\bluetext{C}] Now, follow the edits and make further improvements. As a reference, here is the screenshot of the initial layout: [\redtext{\texttt{I$_{0}$}}] \\
    \bottomrule
  \end{tabular}
  }
  \caption{\small{Construction of multimodal example for each model.
  [\bluetext{\texttt{T}}]: app functionality description.
  [\bluetext{\texttt{C}}]: layout design code snippet.
  [\bluetext{C}]: concatenation of multiple layout design code snippets.
  [\redtext{\texttt{I}}]: visual encoding of image.
  }}
  \label{tab:multimodal_seq}
\end{table}

To unify multimodal features into the same sequence, we use an image tokenizer to generate visual encodings, which are then concatenated with text token embeddings.
For both image and text tokenization, we use the builtin tokenizers in public Gemini-1.5 backbone.

\subsection{Training \& Evaluation} \label{sec:train_eval}

Our experiments focus on applying those model choices discussed in Sec.~\ref{sec:modeling} on top of Gemini-1.5.
Specifically, we use a small version of Gemini with vision understanding capabilities.
For image processing, we use its builtin image tokenizer.

For all the model variants in Sec.~\ref{sec:modeling}, we fine-tune Gemini on our \textsc{Rare+} data for 8k steps, which gives stabilized loss reduction during training and avoids obvious overfitting.
For \emph{Multi-revision} models, we configure the prefix context window to be 16k tokens (including text and image), while using 8k for all other models.
During decoding, we limit the number of tokens to be $400$ and defaults to the temperature of $0.0$ unless otherwise specified.

For evaluation, we adopt the same method as \citep{cheng2023play, nauata2021house}, rendering the layouts as images and computing the FID scores. Due to the limited size of our dataset, we change the sample size for FID from $1024$ to $512$.

\subsection{Direct Inference} \label{sec:direct_inf}
The first question we want to answer is \emph{how our model performs in a direct inference}.
This is essentially evaluating $f_\theta(S_0)$ irrespective of how model was trained and ignoring any revision edits at test time.
For models that rely on revision edits as part of the input, we duplicate $S_0$ since, when sampling training data (e.g., in the \emph{Hop} and \emph{Milti-revision} models), such cases are already covered.
Tab.~\ref{tab:direct_results} shows results in this setting along with the original \textsc{Rare} results as a reference.


We should note that the results between \textsc{Rare} and \textsc{Rare+} are not in strict comparison due to the varying data sizes and FID evaluation sample sizes. Therefore, we fine-tuned PLay, which the original \textsc{Rare} results are based on, on \textsc{Rare+} for fair comparisons. We can observe a significant gap in FID score between \textsc{Rare} and \textsc{Rare+} on PLay-SFT, showing the gain on the new dataset. Moreover, we can also observe a large improvement in FID score between PLay-SFT and other results in the table, highlighting the gain after shifting the backbone model to Gemini, even though our model setups are much simpler.

\begin{table}[h]
  \centering
  \scalebox{1.0}{
  \begin{tabular}{l|c|c|c}
    \toprule
    Data & Model & Config & FID Score ($\downarrow$) \\
    \midrule
    \multirow{2}{*}{\textsc{Rare}} & PLay-SFT & - & 68.9 \\
                                    & PLay-RLHF & - & 68.8 \\
    \midrule
    \midrule
    \multirow{7}{*}{\textsc{Rare+}} & PLay-SFT & $f_\theta(C_0)$ & 31.8 \\
                                    \cmidrule{2-4}
                                    & Direct & $f_\theta(C_0)$ & 24.8 \\
                                    & Direct & $f_\theta(C_i)$ & 23.2 \\
                                    & Hop & j-then-i & 23.1\\
                                    & Hop & quantized & 25.6 \\
                                    & Single revision & - & \textbf{22.3} \\
                                    & Multi-revision & - &  23.2 \\
    \bottomrule
  \end{tabular}
  }
  \caption{\small{Results when evaluating $f_\theta(S_0)$ without revision edits provided at test time.}}
  \label{tab:direct_results}
\end{table}

When comparing different configurations within the same model type, there is a tendency that increasing the input variance leads to better FID scores, suggesting there is a positive correlation between revision edits and the final layout.
For instance, in the \emph{Direct} model, we used $C_i$ to replace $C_0$.
For another instance, \emph{Single/Multi-revision} models perform better then the direct models.
Another interesting observation is that the quantized \emph{Hop} model performed the worst.
This might be due to the supervision target $S_j$ varies too much across quantization buckets.
In short, incorporating edits as the input during training improves performance at test time, even though these edits are not provided at test time, while using these intermediate edits as the training target does not.
Lastly, the \emph{Single revision} model performed the best. In contrast, the \emph{Multi-revision} model did not perform as well as expected, which is likely because it incorporates longer edit context that confuses the model during training and inference.

\subsection{Human Guided Inference} \label{sec:human_guided}

Now, we want to observe how human revision edits impact the performance.
To do that, at test time, we feed the \emph{Single/Multi-revision} models with sampled intermediate states of human revision sequences, and compare the performance of these models against their direct inference.
As shown in Tab.~\ref{tab:multi_results}, incorporating single revision edit improves FID score, and incorporating multiple human edits is even more helpful. Model performance is already close to that of experts (which is around FID $6$).
In Sec.~\ref{sec:copy_paste}\&\ref{sec:quanlitative}, we take a closer look at model behavior in their predictions.

\begin{table}[h]
  \centering
  \scalebox{1.0}{
  \begin{tabular}{l|c|c|c}
    \toprule
    Data & Model & Input Edits & FID Score ($\downarrow$) \\
    \midrule
    \multirow{5}{*}{\textsc{Rare+}} & \multirow{2}{*}{Single revision} & - & 22.3 \\
                                    &  & $S_i$ & 13.7  \\
                                    \cmidrule{2-4}
                                    & \multirow{3}{*}{Multi-revision} & - & 23.2 \\
                                    & & $S_i$ & 17.1 \\
                                    & & $\mathcal{S}$ &  10.5 \\
    \bottomrule
  \end{tabular}
  }
  \caption{\small{Comparison between inference with and without intermediate revision edits of human. \emph{Input Edits}: revision edits used during inference in addition to $S_0$. Human performance is around FID 6.}}
  \label{tab:multi_results}
\end{table}

\section{Analysis}
In this section, we take a closer look at model behavior with layout revision.
In Sec.~\ref{sec:iterative_self_revision}, we check whether our models can iteratively refine design generations.
Sec.~\ref{sec:human_in_the_loop} mixes human and model-generated revisions to examine their gap in terms of performances.
Moreover, Sec.~\ref{sec:copy_paste} analyzes model behavior quantitatively to investigate how often and how much models repeat their previous output.
Lastly, we look at qualitative examples of model generation versus human designs.



\subsection{Iterative Self-revisions} \label{sec:iterative_self_revision}
Prior works such as \citep{blt, nauata2021house, chang2023muse} have demonstrated that iterative inference can improve the quality for UI, floor plan, and image generation, by training on synthetic revisions using randomly masked samples. With the observation that human revision improves layout generation, a natural question arises: \emph{Can model generate self-revisions to iteratively improve its layout design?}
To this end, we chain model outputs as inputs in the next round and conduct such an analysis for 3 iterations.
For the \emph{Direct}/\emph{Hop}, this is basically $S_{i+1}\sim f_\theta(S_i)$ where $i$ starts from 0.
For the \emph{Single revision}, this is $S_{i+1} \sim f_\theta(S_0, S_i)$;
and for the \emph{Multi-revision}, $S_{i+1} \sim f_\theta(S_0, \mathcal{S}_{\leq i})$.
Tab.~\ref{tab:self_revision} shows the FID scores for each round.


\begin{table}[h]
  \centering
  \scalebox{1.0}{
  \begin{tabular}{l|l|c|c|c}
    \toprule
    & Config & Round 1 & Round 2 & Round 3 \\
    \midrule
    Direct & $f_\theta(C_i)$ & 23.2 & 22.7 & 22.6 \\
    \midrule
    \multirow{2}{*}{Hop} & j-then-i & 23.1 & 23.4 & 22.9 \\
        & quantized & 25.6 & 25.1 & 25.0 \\
    \midrule
    Single revision & - & 22.3 & 22.3 & 22.2 \\
    \midrule
    \multirow{3}{*}{Multi-revision} & temp=0.0 & 23.2 & 23.9 & 23.9 \\
        & temp=1.0 & 22.4 & 23.0 & 23.3 \\
        & temp=2.0 & \textbf{18.4} & 18.8 & 19.0 \\
    \bottomrule
  \end{tabular}
  }
  \caption{\small{FID scores $(\downarrow)$ when running models in the chain mode.
  Decoding temperature is $0$ unless otherwise specified.}}
  \label{tab:self_revision}
\end{table}

Again, we start with the decoding temperature equals to $0$.
All of the 4 model classes perform consistently across rounds, neither substantially worse nor better.
Taking a deeper look, a common issue among model generation is that the self-revisions tend to duplicate the input revision, i.e., $S_{i+1} \approx S_i$ irrespective of the model choices.
This often happens after round 1.
To alleviate such an issue, we increased the temperature (from 0 to 2) to induce higher input variance in future rounds (similar to the observation we made in Sec.~\ref{sec:direct_inf}).
Interestingly, doing so did not bring iterative improvement, but instead, increased the round 1 performance by around $4$ points.

As a conclusion, it is difficult to ask models to fully rely on themselves for self-revision.
In Sec.~\ref{sec:copy_paste}, we will show that this is largely due to an echo chamber phenomenon.
Finally, in Sec.~\ref{sec:human_in_the_loop}, we will show how human intervention can help in this case.

\subsection{Iterative Revision with Human in the Loop} \label{sec:human_in_the_loop}

With the observation that model self-revision has limited impact on iterative generation, here, we want to see if the problem can be alleviated in a human-in-the-loop case.
Specifically, we take the \emph{Multi-revision} model and feed it with the a human revision edit during round 1, and then let model to self-revise its generation.
Doing so essentially simulates the scenario where we let human correct model errors at the early stage of the iterative process.
Results are shown in Tab.~\ref{tab:self_revision_with_human}.

\begin{table}[h]
  \centering
  \scalebox{1.0}{
  \begin{tabular}{l|c|c|c}
    \toprule
     & Round 1 & Round 2 & Round 3 \\
    \midrule
    Model $S_i$ & 23.2 & 23.9 & 23.9 \\
    Human $S_i$ & 17.1 & 18.0 & 18.3 \\
    Human $\mathcal{S}$ & 10.5 & - & - \\
    \bottomrule
  \end{tabular}
  }
  \caption{\small{Iterative generation using different intermediate edits. Model denotes the \emph{Multi-revision}.}}
  \label{tab:self_revision_with_human}
\end{table}

There is a clear gap between using human edits and model-generated ones in the first round.
Human edits turn out to be effective as it rapidly improved the FID score from $23.2$ to $17.1$.
Consistent with Tab.~\ref{tab:self_revision}, self-revision still does not bring iterative improvement.
In this analysis, the practical upper bound is $10.5$ since it uses a set of revision edits from human.
Closing the gap from $17.1$ to $10.5$ requires aligning human edits and model generated ones.
We leave this challenge to future works.

\subsection{Do Our Models Copy-paste or Actually Edit?} \label{sec:copy_paste}

In Sec.~\ref{sec:iterative_self_revision}, we have seen that model's self-revisions tend to copy its input.
Here, we want to know if models are mostly copy-pasting the previous generated revisions or actually doing editing.
We answer this question via a comprehensive analysis to see how often and how much that happens.
Specifically, we use two metrics.
One is the percentage of design code that is identical to the previous round (i.e., $C_{i+1} = C_i$).
We refer to this measurement as $\rho$ in Tab.~\ref{tab:output_stats}.
Another metric is the standard \textsc{RougeL}~\citep{lin-2004-rouge} where we use $C_i$ as reference and $C_{i+1}$ as hypothesis for round $i+1$.

\begin{table}[h]
  \centering
  \scalebox{1.0}{
  \begin{tabular}{l|l|c|c|c|c|c|c}
    \toprule
    & Config & \multicolumn{2}{|c|}{Round 1} & \multicolumn{2}{|c|}{Round 2} & \multicolumn{2}{|c}{Round 3} \\
    \midrule
    & & $\rho$ & \textsc{RougeL} & $\rho$ & \textsc{RougeL} & $\rho$ & \textsc{RougeL} \\
    \midrule
    Direct & $f_\theta(C_i)$ & 3.4 & 36.9 & 61.1 & 90.6 & 91.0 & 97.6 \\
    \midrule
    \multirow{2}{*}{Hop} & j-then-i & 0.3 & 32.5 & 41.0 & 87.5 & 72.0 & 94.6 \\
        & quantized & 1.1 & 37.2 & 44.5 & 88.2 & 77.9 & 95.4 \\
    \midrule
    Single revision & - & 3.0 & 32.9 & 76.3 & 96.6 & 93.4 & 99.0  \\
    \midrule
    \multirow{3}{*}{Multi-revision} & temp=0.0 & 0.5 & 30.6 & 64.6 & 96.2 & 81.3 & 98.8 \\
        & temp=1.0 & 0.5 & 30.2 & 57.0 & 93.9 & 71.6 & 97.3 \\
        & temp=2.0 & 0.5 & 29.8 & 24.0 & 83.1 & 26.4 & 85.5 \\
    \bottomrule
  \end{tabular}
  }
  \caption{\small{Output statistics of model predictions in Tab.~\ref{tab:self_revision}.
  $\rho$: percent of [$C_i = C_{i+1}$];
  \textsc{RougeL}: textual metric between $C_i$ and $C_{i+1}$.}}
  \label{tab:output_stats}
\end{table}

Clearly, the models perform editing to the input layout design.
That is, a low percentage of generations ($\rho$) that are identical to the input ones, as well as a reasonable \textsc{RougeL} among the two.
However, after round 1, the models tend to repeat themselves.
This is observed in the sharp increase in $\rho$ and close to $100$ \textsc{RougeL} scores.
Furthermore, we see that increasing the decoding temperature remedies this phenomenon (i.e., reducing $\rho$ to 20-ish and \textsc{RougeL} to 80-ish) but fundamentally does not avoid repeating the trend.
While addressing such an echo chamber issue is not the focus of this paper, we demonstrate the severity of the issue and the potential of solving the problem.

\subsection{Qualitative Examples of Edits} \label{sec:quanlitative}

\begin{figure*}[h]
  \centering
   \includegraphics[width=0.95\linewidth]{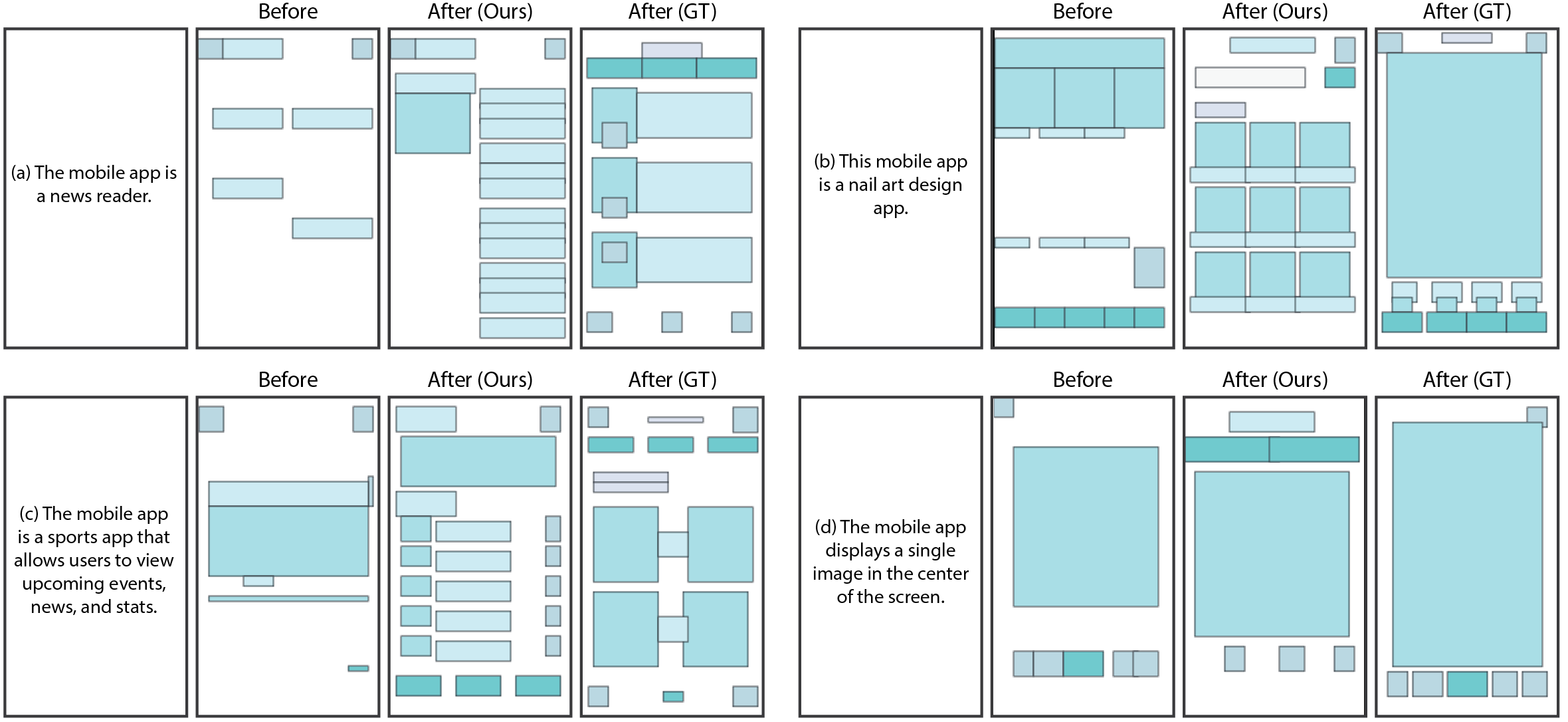}

   \caption{Qualitative analysis of model generations vs ground truths.
   See Appx.~\ref{sec:color_legend} for color legends.}
   \label{fig:quanlitative_examples}
\end{figure*}

In this section, we look at example of model generation in comparison with the starting state and the ground truth.
Fig.~\ref{fig:quanlitative_examples} shows that our model generation carry relatively more visual patterns from the input state, such as repeating and extending a given pattern more times.
In contrast, human edits (GT) perform more drastic design changes.
Patterns that present in the original state are often removed or replaced by others.
In Appx.~\ref{sec:generation_examples}, we show more generated examples.

\section{Conclusions}
In this paper, we have demonstrated the significant value of incorporating human revision edits into multimodal generative models for layout design. By utilizing the newly collected \textsc{Rare+} dataset, which captures the iterative design process, supervised fine-tuning setups were explored, leading to substantial improvements over models that ignored these edits. Notably, the best model variant achieved a remarkable FID score of approximately 10, closely approaching human performance. Despite the inherent noise in human edits, we highlight the critical role of human guidance in preventing self-repeating behavior in auto-regressive generation. Overall, these findings emphasize the potential of harnessing human revisions to enhance the quality and align with the iterative nature of layout design.

\section*{Limitations}
In this paper, we explored various modeling setups based on supervised fine-tuning Gemini.
While the noisy and lengthy edit sequences makes it difficult to formulate reward learning, we believe seeing it with a different angle from the instruction tuning works could make the data using more sophisticated.
For instance, revision edits can be heuristically converted to natural language instructions.
Such edits can potentially be combined with in-context learning for layout generation~\cite{lin2024layoutprompter} with a richer form of task command $T$.
How such reformulation of the data can benefit design generation awaits to be seen.
Moreover, we focused on modeling the final output generation in this paper
Future work could benefit from the large memory capacity in large multimodal backbone to model the entire revision trajectory (both states and actions) in a fully auto-regressive manner.
Lastly, we worked around the echo chamber issue by injecting human revision edit at an early stage of the iterative process.
How can we automatically correct such error plays critical factor. However, such issue is not addressed in this work.

\section*{Broader Impact}
Looking forward, the integration of automatic design tools for layout design process could further enhance the efficiency and quality of design-related tasks. The principles and findings from our approach provide a robust foundation for future research in human-in-the-loop models, particularly in aspects of design and creativity tasks. As AI continues to integrate into diverse aspects of decision-making, embedding introspective capabilities will be essential to ensure these systems operate not only with precision but with an understanding akin to strategic human cognition.

\section*{Ethics Statement}
As the capabilities of large models increase, it is crucial to address potential ethical concerns, particularly regarding data privacy, bias, and transparency. Our work focuses on leveraging human revisions to improve generative models for layout design, aiming to enable human-in-the-loop for creative work. We emphasize the importance of human revision to monitor and mitigate unforeseen consequences and encourage the responsible use of this technology for societal benefit. 

\bibliography{citations}

\begin{thebibliography}{33}
\expandafter\ifx\csname natexlab\endcsname\relax\def\natexlab#1{#1}\fi

\bibitem[{Adeniji et~al.(2023)Adeniji, Xie, Sferrazza, Seo, James, and
  Abbeel}]{adeniji2023language}
Ademi Adeniji, Amber Xie, Carmelo Sferrazza, Younggyo Seo, Stephen James, and
  Pieter Abbeel. 2023.
\newblock \href {http://arxiv.org/abs/2308.12270} {Language Reward Modulation
  for Pretraining Reinforcement Learning}.

\bibitem[{Arroyo et~al.(2021)Arroyo, Postels, and Tombari}]{vtn}
Diego~Martin Arroyo, Janis Postels, and Federico Tombari. 2021.
\newblock Variational transformer networks for layout generation.
\newblock In \emph{Proceedings of the IEEE/CVF Conference on Computer Vision
  and Pattern Recognition}, pages 13642--13652.

\bibitem[{Bai et~al.(2022)Bai, Jones, Ndousse, Askell, Chen, DasSarma, Drain,
  Fort, Ganguli, Henighan, Joseph, Kadavath, Kernion, Conerly, El-Showk,
  Elhage, Hatfield-Dodds, Hernandez, Hume, Johnston, Kravec, Lovitt, Nanda,
  Olsson, Amodei, Brown, Clark, McCandlish, Olah, Mann, and
  Kaplan}]{bai2022training}
Yuntao Bai, Andy Jones, Kamal Ndousse, Amanda Askell, Anna Chen, Nova DasSarma,
  Dawn Drain, Stanislav Fort, Deep Ganguli, Tom Henighan, Nicholas Joseph,
  Saurav Kadavath, Jackson Kernion, Tom Conerly, Sheer El-Showk, Nelson Elhage,
  Zac Hatfield-Dodds, Danny Hernandez, Tristan Hume, Scott Johnston, Shauna
  Kravec, Liane Lovitt, Neel Nanda, Catherine Olsson, Dario Amodei, Tom Brown,
  Jack Clark, Sam McCandlish, Chris Olah, Ben Mann, and Jared Kaplan. 2022.
\newblock \href {http://arxiv.org/abs/2204.05862} {Training a Helpful and
  Harmless Assistant with Reinforcement Learning from Human Feedback}.

\bibitem[{Black et~al.(2023)Black, Janner, Du, Kostrikov, and
  Levine}]{black2023training}
Kevin Black, Michael Janner, Yilun Du, Ilya Kostrikov, and Sergey Levine. 2023.
\newblock \href {http://arxiv.org/abs/2305.13301} {Training Diffusion Models
  with Reinforcement Learning}.

\bibitem[{Casper et~al.(2023)Casper, Davies, Shi, Gilbert, Scheurer, Rando,
  Freedman, Korbak, Lindner, Freire, Wang, Marks, Segerie, Carroll, Peng,
  Christoffersen, Damani, Slocum, Anwar, Siththaranjan, Nadeau, Michaud, Pfau,
  Krasheninnikov, Chen, Langosco, Hase, Bıyık, Dragan, Krueger, Sadigh, and
  Hadfield-Menell}]{casper2023open}
Stephen Casper, Xander Davies, Claudia Shi, Thomas~Krendl Gilbert, Jérémy
  Scheurer, Javier Rando, Rachel Freedman, Tomasz Korbak, David Lindner, Pedro
  Freire, Tony Wang, Samuel Marks, Charbel-Raphaël Segerie, Micah Carroll,
  Andi Peng, Phillip Christoffersen, Mehul Damani, Stewart Slocum, Usman Anwar,
  Anand Siththaranjan, Max Nadeau, Eric~J. Michaud, Jacob Pfau, Dmitrii
  Krasheninnikov, Xin Chen, Lauro Langosco, Peter Hase, Erdem Bıyık, Anca
  Dragan, David Krueger, Dorsa Sadigh, and Dylan Hadfield-Menell. 2023.
\newblock \href {http://arxiv.org/abs/2307.15217} {Open Problems and
  Fundamental Limitations of Reinforcement Learning from Human Feedback}.

\bibitem[{Chang et~al.(2023)Chang, Zhang, Barber, Maschinot, Lezama, Jiang,
  Yang, Murphy, Freeman, Rubinstein et~al.}]{chang2023muse}
Huiwen Chang, Han Zhang, Jarred Barber, AJ~Maschinot, Jose Lezama, Lu~Jiang,
  Ming-Hsuan Yang, Kevin Murphy, William~T Freeman, Michael Rubinstein, et~al.
  2023.
\newblock Muse: Text-to-image generation via masked generative transformers.
\newblock \emph{arXiv preprint arXiv:2301.00704}.

\bibitem[{Chen et~al.(2022)Chen, Wang, Changpinyo, Piergiovanni, Padlewski,
  Salz, Goodman, Grycner, Mustafa, Beyer et~al.}]{chen2022pali}
Xi~Chen, Xiao Wang, Soravit Changpinyo, AJ~Piergiovanni, Piotr Padlewski,
  Daniel Salz, Sebastian Goodman, Adam Grycner, Basil Mustafa, Lucas Beyer,
  et~al. 2022.
\newblock Pali: A jointly-scaled multilingual language-image model.
\newblock \emph{arXiv preprint arXiv:2209.06794}.

\bibitem[{Cheng et~al.(2023)Cheng, Huang, Li, and Li}]{cheng2023play}
Chin-Yi Cheng, Forrest Huang, Gang Li, and Yang Li. 2023.
\newblock \href {http://arxiv.org/abs/2301.11529} {PLay: Parametrically
  Conditioned Layout Generation using Latent Diffusion}.

\bibitem[{Fan et~al.(2023)Fan, Watkins, Du, Liu, Ryu, Boutilier, Abbeel,
  Ghavamzadeh, Lee, and Lee}]{fan2023dpok}
Ying Fan, Olivia Watkins, Yuqing Du, Hao Liu, Moonkyung Ryu, Craig Boutilier,
  Pieter Abbeel, Mohammad Ghavamzadeh, Kangwook Lee, and Kimin Lee. 2023.
\newblock \href {http://arxiv.org/abs/2305.16381} {DPOK: Reinforcement Learning
  for Fine-tuning Text-to-Image Diffusion Models}.

\bibitem[{Gupta et~al.(2021)Gupta, Lazarow, Achille, Davis, Mahadevan, and
  Shrivastava}]{gupta2021layouttransformer}
Kamal Gupta, Justin Lazarow, Alessandro Achille, Larry~S Davis, Vijay
  Mahadevan, and Abhinav Shrivastava. 2021.
\newblock Layouttransformer: Layout generation and completion with
  self-attention.
\newblock In \emph{Proceedings of the IEEE/CVF International Conference on
  Computer Vision}, pages 1004--1014.

\bibitem[{Hejna and Sadigh(2022)}]{hejna2022fewshot}
Joey Hejna and Dorsa Sadigh. 2022.
\newblock \href {http://arxiv.org/abs/2212.03363} {Few-Shot Preference Learning
  for Human-in-the-Loop RL}.

\bibitem[{Heusel et~al.(2017)Heusel, Ramsauer, Unterthiner, Nessler, and
  Hochreiter}]{heusel2017gans}
Martin Heusel, Hubert Ramsauer, Thomas Unterthiner, Bernhard Nessler, and Sepp
  Hochreiter. 2017.
\newblock Gans trained by a two time-scale update rule converge to a local nash
  equilibrium.
\newblock \emph{Advances in neural information processing systems}, 30.

\bibitem[{Inoue et~al.(2023)Inoue, Kikuchi, Simo-Serra, Otani, and
  Yamaguchi}]{inoue2023layoutdm}
Naoto Inoue, Kotaro Kikuchi, Edgar Simo-Serra, Mayu Otani, and Kota Yamaguchi.
  2023.
\newblock LayoutDM: Discrete Diffusion Model for Controllable Layout
  Generation.
\newblock In \emph{Proceedings of the IEEE/CVF Conference on Computer Vision
  and Pattern Recognition}, pages 10167--10176.

\bibitem[{Jain et~al.(2022)Jain, Xie, and Abbeel}]{jain2022vectorfusion}
Ajay Jain, Amber Xie, and Pieter Abbeel. 2022.
\newblock VectorFusion: Text-to-SVG by Abstracting Pixel-Based Diffusion
  Models.
\newblock \emph{arXiv}.

\bibitem[{Kikuchi et~al.(2021)Kikuchi, Simo-Serra, Otani, and
  Yamaguchi}]{kikuchi2021constrained}
Kotaro Kikuchi, Edgar Simo-Serra, Mayu Otani, and Kota Yamaguchi. 2021.
\newblock Constrained graphic layout generation via latent optimization.
\newblock In \emph{Proceedings of the 29th ACM International Conference on
  Multimedia}, pages 88--96.

\bibitem[{Kong et~al.(2022)Kong, Jiang, Chang, Zhang, Hao, Gong, and
  Essa}]{blt}
Xiang Kong, Lu~Jiang, Huiwen Chang, Han Zhang, Yuan Hao, Haifeng Gong, and
  Irfan Essa. 2022.
\newblock BLT: bidirectional layout transformer for controllable layout
  generation.
\newblock In \emph{Computer Vision--ECCV 2022: 17th European Conference, Tel
  Aviv, Israel, October 23--27, 2022, Proceedings, Part XVII}, pages 474--490.
  Springer.

\bibitem[{Lee et~al.(2023)Lee, Liu, Ryu, Watkins, Du, Boutilier, Abbeel,
  Ghavamzadeh, and Gu}]{lee2023aligning}
Kimin Lee, Hao Liu, Moonkyung Ryu, Olivia Watkins, Yuqing Du, Craig Boutilier,
  Pieter Abbeel, Mohammad Ghavamzadeh, and Shixiang~Shane Gu. 2023.
\newblock \href {http://arxiv.org/abs/2302.12192} {Aligning Text-to-Image
  Models using Human Feedback}.

\bibitem[{Lee et~al.(2021)Lee, Smith, and Abbeel}]{lee2021pebble}
Kimin Lee, Laura Smith, and Pieter Abbeel. 2021.
\newblock Pebble: Feedback-efficient interactive reinforcement learning via
  relabeling experience and unsupervised pre-training.
\newblock \emph{arXiv preprint arXiv:2106.05091}.

\bibitem[{Lin(2004)}]{lin-2004-rouge}
Chin-Yew Lin. 2004.
\newblock \href {https://aclanthology.org/W04-1013} {{ROUGE}: A Package for
  Automatic Evaluation of Summaries}.
\newblock In \emph{Text Summarization Branches Out}, pages 74--81, Barcelona,
  Spain. Association for Computational Linguistics.

\bibitem[{Lin et~al.(2023)Lin, Guo, Sun, Xu, Liu, Lou, and
  Zhang}]{lin2023parse}
Jiawei Lin, Jiaqi Guo, Shizhao Sun, Weijiang Xu, Ting Liu, Jian-Guang Lou, and
  Dongmei Zhang. 2023.
\newblock A parse-then-place approach for generating graphic layouts from
  textual descriptions.
\newblock In \emph{Proceedings of the IEEE/CVF International Conference on
  Computer Vision}, pages 23622--23631.

\bibitem[{Lin et~al.(2024)Lin, Guo, Sun, Yang, Lou, and
  Zhang}]{lin2024layoutprompter}
Jiawei Lin, Jiaqi Guo, Shizhao Sun, Zijiang Yang, Jian-Guang Lou, and Dongmei
  Zhang. 2024.
\newblock LayoutPrompter: Awaken the Design Ability of Large Language Models.
\newblock \emph{Advances in Neural Information Processing Systems}, 36.

\bibitem[{Liu et~al.(2023)Liu, Sferrazza, and Abbeel}]{liu2023chain}
Hao Liu, Carmelo Sferrazza, and Pieter Abbeel. 2023.
\newblock \href {http://arxiv.org/abs/2302.02676} {Chain of Hindsight Aligns
  Language Models with Feedback}.

\bibitem[{Nair et~al.(2022)Nair, Rajeswaran, Kumar, Finn, and
  Gupta}]{nair2022r3m}
Suraj Nair, Aravind Rajeswaran, Vikash Kumar, Chelsea Finn, and Abhinav Gupta.
  2022.
\newblock \href {http://arxiv.org/abs/2203.12601} {R3M: A Universal Visual
  Representation for Robot Manipulation}.

\bibitem[{Nauata et~al.(2021)Nauata, Hosseini, Chang, Chu, Cheng, and
  Furukawa}]{nauata2021house}
Nelson Nauata, Sepidehsadat Hosseini, Kai-Hung Chang, Hang Chu, Chin-Yi Cheng,
  and Yasutaka Furukawa. 2021.
\newblock House-gan++: Generative adversarial layout refinement network towards
  intelligent computational agent for professional architects.
\newblock In \emph{Proceedings of the IEEE/CVF Conference on Computer Vision
  and Pattern Recognition}, pages 13632--13641.

\bibitem[{Ouyang et~al.(2022)Ouyang, Wu, Jiang, Almeida, Wainwright, Mishkin,
  Zhang, Agarwal, Slama, Ray, Schulman, Hilton, Kelton, Miller, Simens, Askell,
  Welinder, Christiano, Leike, and Lowe}]{ouyang2022training}
Long Ouyang, Jeff Wu, Xu~Jiang, Diogo Almeida, Carroll~L. Wainwright, Pamela
  Mishkin, Chong Zhang, Sandhini Agarwal, Katarina Slama, Alex Ray, John
  Schulman, Jacob Hilton, Fraser Kelton, Luke Miller, Maddie Simens, Amanda
  Askell, Peter Welinder, Paul Christiano, Jan Leike, and Ryan Lowe. 2022.
\newblock \href {http://arxiv.org/abs/2203.02155} {Training language models to
  follow instructions with human feedback}.

\bibitem[{Rafailov et~al.(2023)Rafailov, Sharma, Mitchell, Ermon, Manning, and
  Finn}]{rafailov2023direct}
Rafael Rafailov, Archit Sharma, Eric Mitchell, Stefano Ermon, Christopher~D.
  Manning, and Chelsea Finn. 2023.
\newblock \href {http://arxiv.org/abs/2305.18290} {Direct Preference
  Optimization: Your Language Model is Secretly a Reward Model}.

\bibitem[{Rombach et~al.(2022)Rombach, Blattmann, Lorenz, Esser, and
  Ommer}]{rombach2022highresolution}
Robin Rombach, Andreas Blattmann, Dominik Lorenz, Patrick Esser, and Björn
  Ommer. 2022.
\newblock \href {http://arxiv.org/abs/2112.10752} {High-Resolution Image
  Synthesis with Latent Diffusion Models}.

\bibitem[{Sharma et~al.(2024)Sharma, Liao, and Xiao}]{sharma2024generative}
Nikhil Sharma, Q~Vera Liao, and Ziang Xiao. 2024.
\newblock Generative Echo Chamber? Effect of LLM-Powered Search Systems on
  Diverse Information Seeking.
\newblock In \emph{Proceedings of the CHI Conference on Human Factors in
  Computing Systems}, pages 1--17.

\bibitem[{Stiennon et~al.(2022)Stiennon, Ouyang, Wu, Ziegler, Lowe, Voss,
  Radford, Amodei, and Christiano}]{stiennon2022learning}
Nisan Stiennon, Long Ouyang, Jeff Wu, Daniel~M. Ziegler, Ryan Lowe, Chelsea
  Voss, Alec Radford, Dario Amodei, and Paul Christiano. 2022.
\newblock \href {http://arxiv.org/abs/2009.01325} {Learning to summarize from
  human feedback}.

\bibitem[{Team et~al.(2023)Team, Anil, Borgeaud, Wu, Alayrac, Yu, Soricut,
  Schalkwyk, Dai, Hauth et~al.}]{team2023gemini}
Gemini Team, Rohan Anil, Sebastian Borgeaud, Yonghui Wu, Jean-Baptiste Alayrac,
  Jiahui Yu, Radu Soricut, Johan Schalkwyk, Andrew~M Dai, Anja Hauth, et~al.
  2023.
\newblock Gemini: a family of highly capable multimodal models.
\newblock \emph{arXiv preprint arXiv:2312.11805}.

\bibitem[{Vaswani et~al.(2017)Vaswani, Shazeer, Parmar, Uszkoreit, Jones,
  Gomez, Kaiser, and Polosukhin}]{transformer}
Ashish Vaswani, Noam Shazeer, Niki Parmar, Jakob Uszkoreit, Llion Jones,
  Aidan~N Gomez, {\L}ukasz Kaiser, and Illia Polosukhin. 2017.
\newblock Attention is all you need.
\newblock \emph{Advances in neural information processing systems}, 30.

\bibitem[{Xie et~al.(2024)Xie, Cheng, Huang, and Li}]{xie2024leveraging}
Amber Xie, Chin-Yi Cheng, Forrest Huang, and Yang Li. 2024.
\newblock \href {https://openreview.net/forum?id=3bmjHYX42n} {Leveraging Human
  Revisions for Improving Text-to-Layout Models}.

\bibitem[{Xu et~al.(2023)Xu, Liu, Wu, Tong, Li, Ding, Tang, and
  Dong}]{xu2023imagereward}
Jiazheng Xu, Xiao Liu, Yuchen Wu, Yuxuan Tong, Qinkai Li, Ming Ding, Jie Tang,
  and Yuxiao Dong. 2023.
\newblock \href {http://arxiv.org/abs/2304.05977} {ImageReward: Learning and
  Evaluating Human Preferences for Text-to-Image Generation}.

\end{thebibliography}
\bibliographystyle{acl_natbib}

\newpage
\appendix
\onecolumn
\section{Appendix}

\subsection{Color Legend} \label{sec:color_legend}
\begin{figure}[h!]
    \centering
    \includegraphics[width=0.5\textwidth]{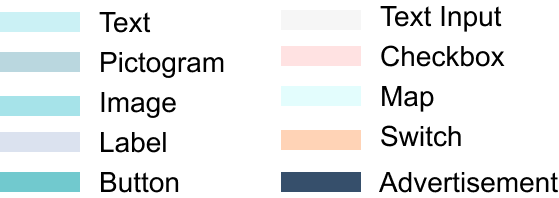}
    \caption{\textbf{Visualization Colors}}
    \label{fig:play_colors}
\end{figure}

We use the same color legend as in \citet{cheng2023play} to visualize the layouts. Colors for popular class elements are rendered in Figure~\ref{fig:play_colors}.

\subsection{Generation Examples} \label{sec:generation_examples}

\begin{figure}[h!]
    \centering
    \includegraphics[width=0.5\textwidth]{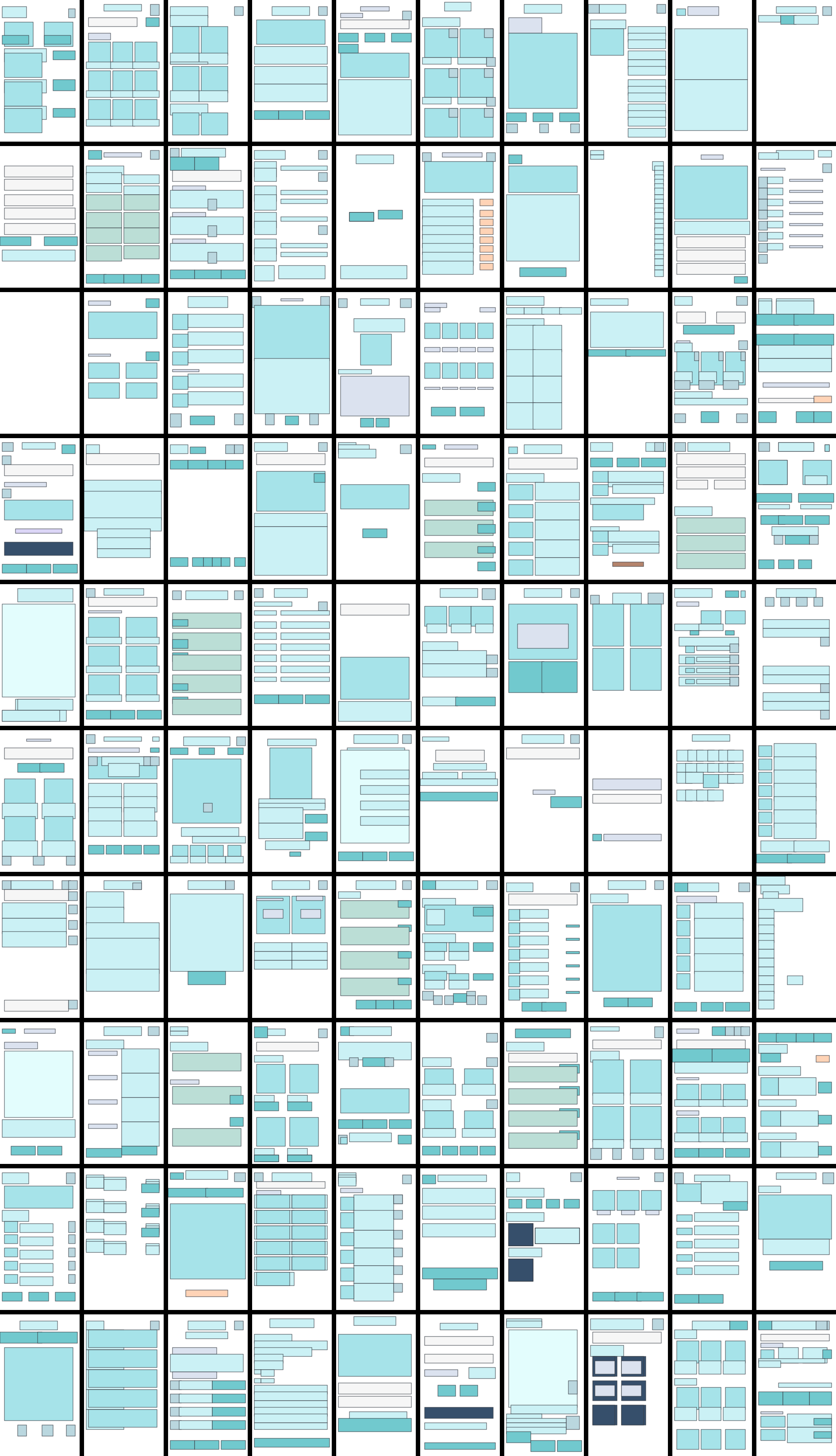}
    \caption{\textbf{Visualization Colors}}
    \label{fig:generation_examples}
\end{figure}

Fig.~\ref{fig:generation_examples} shows more generated examples by our \emph{Multi-revisions} model.
We present them as-is without extra filtering.
Besides a few corner cases where the generated code violates formating requirement (e.g., negative coordinates), model generations are quite consistent in quality (e.g., alignedness and space saturation).

\end{document}